# How We Learn About our Networked World

Sophia U. David[1], Sophie E. Loman[1], Christopher W. Lynn[1,2], Ann S. Blevins[1], Danielle S. Bassett[1-6]

[1]Department of Bioengineering, School of Engineering & Applied Science, University of Pennsylvania, Philadelphia, PA 19104 USA
[2]Department of Physics & Astronomy, College of Arts & Sciences University of Pennsylvania, Philadelphia, PA 19104 USA
[3]Department of Electrical & Systems Engineering, University of Pennsylvania, Philadelphia, PA 19104 USA
[4]Department of Neurology, Perelman School of Medicine, University of Pennsylvania, Philadelphia, PA 19104 USA
[5]Department of Psychiatry, Perelman School of Medicine, University of Pennsylvania, Philadelphia, PA 19104 USA
[6]Santa Fe Institute, Santa Fe, NM 87501 USA
[7]To whom correspondence should be addressed: dsb@seas.upenn.edu

**Abstract**

When presented with information of any type, from music to language to mathematics, the human mind subconsciously arranges it into a network. A network puts pieces of information like musical notes, syllables, or mathematical concepts into context by linking them together. These networks help our minds organize information and anticipate what is coming. Here we present two questions about network building. 1) Can humans more easily learn some types of networks than others? 2) Do humans find some links between ideas more surprising than others? The answer to both questions is "Yes," and we explain why. The findings provide much-needed insight into the ways that humans learn about the networked world around them. Moreover, the study paves the way for future efforts seeking to optimize how information is presented to accelerate human learning.

**Keywords**: Network, learning, mind, cognitive science

## What is a network?

Whether we realize it or not, our brains are constantly making predictions about what will happen next. If you see lightning, you might predict that you will hear thunder. If you see the letter A, you might expect that the letter B will follow. If your dog barks, you might assume that a stranger has arrived at your door. Taking into account the transitions between past events, our brain anticipates future events. Our ability to make these predictions relies on networks or webs of knowledge built from observations and the transitions between them[1].

Networks are made of **nodes** (things) and **edges** (relations among things); see Fig. 1a., *left*. In the thunder and lightning example, the nodes are the thunder and lightning. They are connected by an edge, which represents a possible transition between them. Beyond predicting the future, networks can have all sorts of other functions. They exist among characters in a story, follower and followee connections on Instagram, and syllables in language. The human brain is not only able to interpret and learn networks like these, but also to build its own network model of the world[2].

Why are networks so important? Networks put the interconnected world in a context that we can quantify and measure. For example, the number of connections held by a node is its **degree** (Fig. 1a., *middle*). When you smell smoke, you might see a fire or a barbeque. The smoke node, therefore, has two edges and a degree of two. Each edge, such as would link lightning and thunder, represents a possible **transition,** or movement between linked nodes, that can occur **(**Fig. 1a., *right*). Together, the connections between nodes create a **network structure.** Here, we study three types of network structures (Fig. 1b.): (i) a modular network, here with three groups or clusters, (ii) a lattice network with a repeated structure, and (iii) a ring network in which nodes connect to those close to them in a circle.

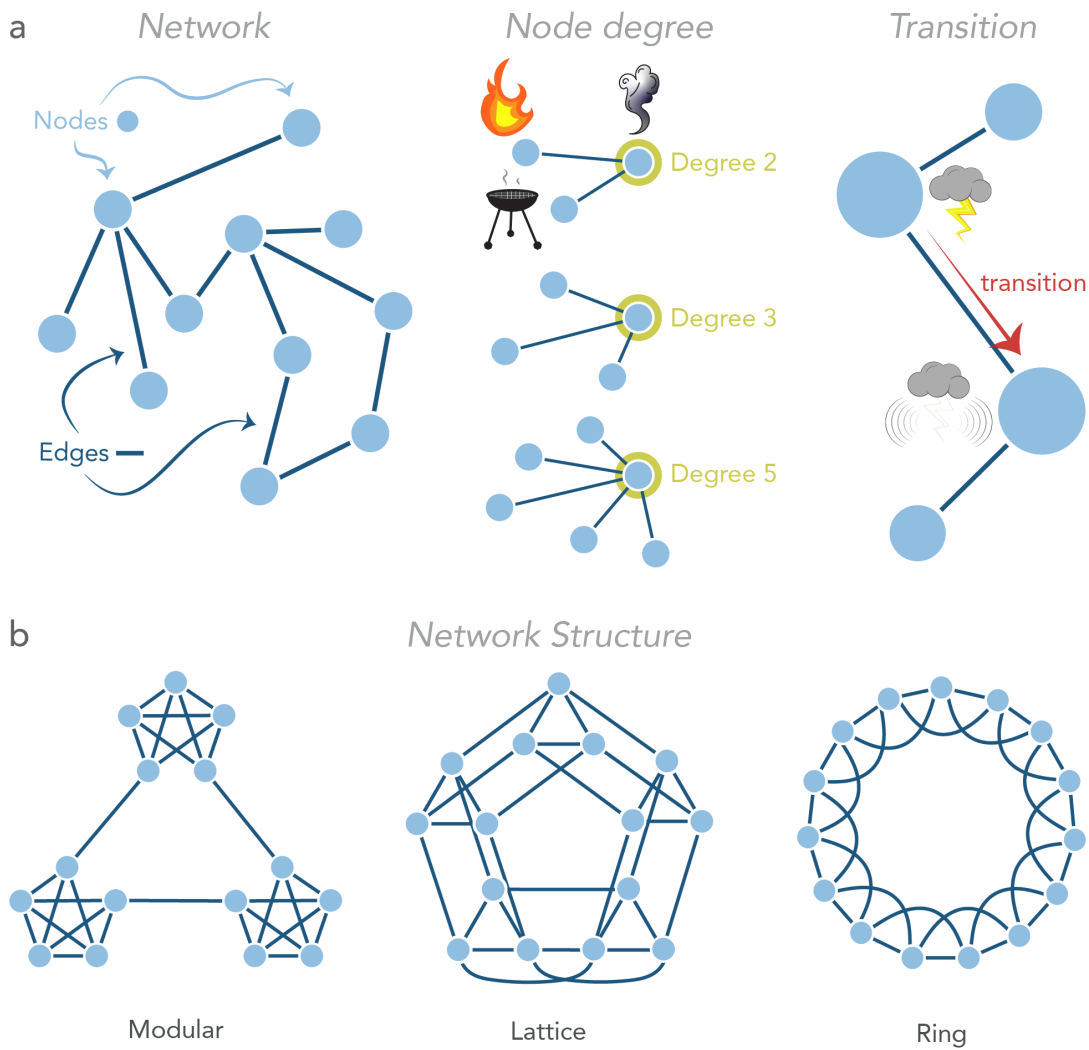

**Figure 1. Understanding how we model our world requires understanding networks and their structure.** (a) Networks can encapsulate predictions. (*left*) A network is a web of connected nodes and edges. (*middle*) Some nodes have more edges than others, or higher *degrees*. (*right*) A transition from one node to another can occur between any pair of connected nodes. (b) Network structures: (*left*) A modular network has dense clusters of interconnected nodes. (*middle*) A lattice network has evenly distributed nodes and edges. (*right*) A ring network is a circle of nodes; each node is connected to its neighboring and next-neighboring nodes.

### Do humans learn some networks better than others?

When we are exposed to new information, we organize it into a network. This representation allows us to specify the connections between pieces of information and to anticipate future events. But, when given a premade network to learn, are humans sensitive to the shape and structure of a network? Can network structure affect how humans learn? To find out, researchers created a network of fifteen images, where each image was randomly assigned to a node in a network, each with four edges (Fig. 2)[3]. The 15 images each contained five gray squares corresponding to the five fingers of the right hand, with squares highlighted (it turns out there are precisely 15 different ways that a human can press 1 or 2 keys on a keyboard, using only 5 keys, with one finger on each -- try it!). When an image appeared, subjects pressed keys that corresponded to the highlighted squares, similar to pressing keys corresponding to notes in guitar hero. Immediately after pressing a key, the image would change to a new image that was connected to the previous image in the network.

Importantly, subjects were never actually shown the network of connections between images (Fig. 2). Instead, subjects were shown one image at a time and had to learn the network of transitions by pure observation. Is there a movie that you have watched so many times that you know exactly what each character will say and when? Similarly, by repetitively observing transitions, subjects slowly learned to predict which image was coming next. The researchers could measure how well a subject was learning the network by recording the time between an image appearing and the subject pressing the corresponding keys. A quick reaction indicated that the subject anticipated the transition, based on their understanding of the network structure, whereas a slow reaction indicated that the subject was surprised by the transition, and therefore had a worse understanding of the network structure (Fig. 2). This experiment provided two key insights regarding how humans learn networks.

1) *Reaction times in the network with **modular structure** were smaller than those in the network with lattice structure* (Fig. 2). Whereas a modular network structure has multiple dense clusters of nodes, a lattice network structure has no clusters and the nodes are spread out evenly (Fig. 1). Even though both modular and lattice structures only contained nodes of the same degree, subjects had an easier time anticipating transitions between nodes in the modular network than in the lattice network (Fig. 2). This observation suggests that humans learn and understand connections between people, objects, and events better when they are categorized into small groups or modules. But how does this behavior relate to networks? The answer lies in how subjects reacted to transitions within clusters versus transitions between clusters in the modular network. Subjects responded much faster to transitions within the same cluster, indicating that the clusters themselves help humans learn the structure of the network and anticipate future events.

2) *Reaction times decreased when probabilities of transitions increased*. In other words, when subjects knew that a transition was likely, they were more easily able to anticipate the future. Regardless of network structure, people more quickly anticipated transitions following nodes of low degree than transitions following nodes of high degree. Intuitively, high degree nodes have more possible connections than low degree nodes, making it more difficult to predict the node to which they will transition. For example, it might be difficult to predict how a baby will react to an event because their prior responses have included crying, laughing, and throwing things. However, it is easy to predict a person's response to their favorite meal because their prior responses have always been simply happiness. This finding suggests that humans are able to pick up on network structure and relationships between nodes in the immediate vicinity of each other.

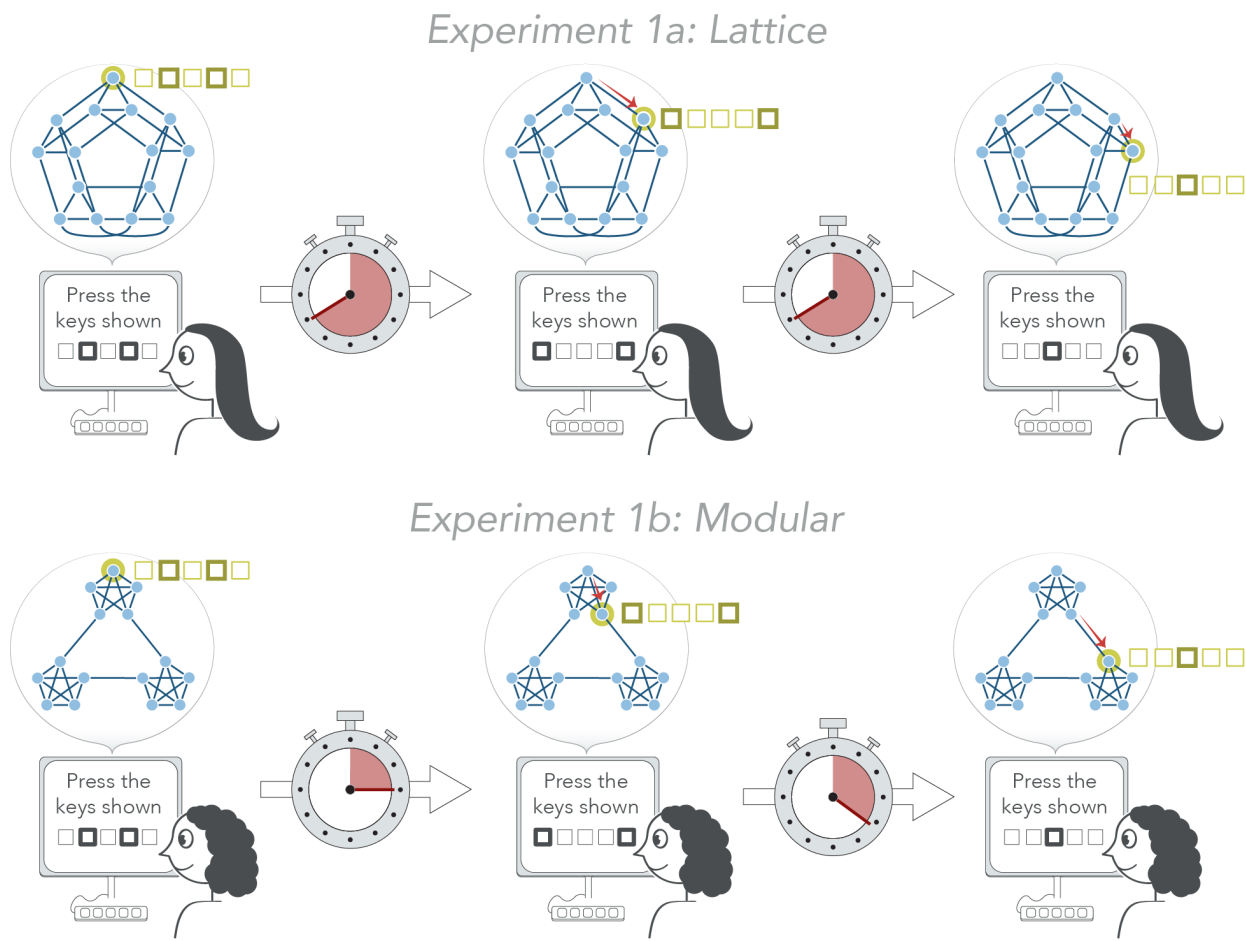

**Figure 2. Building a laboratory to study how humans learn networks.** Subjects were shown 5 squares on a computer screen, with one or two squares highlighted. They used a keyboard to press keys that correspond to the squares highlighted on the screen. After pressing the correct keys, the screen would show a new image or node that is connected to the previous node in the underlying network. (a) When shown information arranged in a lattice network structure, subjects took longer to respond than (b) when shown information arranged in a modular network structure.

### Are some surprises more surprising than others?

We just described how the human brain can learn the structure of a static network simply by observing transitions between nodes, but in real life networks often change in unexpected ways. Because our world is constantly changing, our brain must find new connections between previously unconnected ideas, events, and people[4]. What happens when humans are given new information that does not fit into their learned network at all? Imagine reading the end of a mystery novel, which reveals that the character you thought was good is actually the villain. You might feel surprised because this does not fit into your previous understanding of the connections between the characters.

Once humans learn a network, they come to expect transitions that they have seen before, and when a new transition appears (that is, a transition that "violates" their learned network), humans are surprised. Some network-violating transitions, however, surprise humans more than others. In a new experiment, subjects were asked to respond to 1500 transitions between pieces of information arranged in a **ring structure** (Fig. 1, bottom right)[3]. Among the 1500 transitions shown to subjects, 50 were violations of the network. In these instances, one node transitioned to another even though they were not connected in the network. There were two different types of violations: short violations, in which nodes transitioned to other nodes that were two edges apart, and long violations, in which nodes transitioned to nodes that were three or four edges apart (Fig. 3). As in the previous experiment, researchers recorded subjects' reaction times to measure how surprised they were by the different network violations.

      One might assume that people are equally surprised by all types of network violations, in which case subjects would respond equally quickly to both short and long violations. However, this was not the case. While subjects responded more slowly for both types of violations than standard transitions, their reaction times were even longer for long violations than short violations. This means that humans are less surprised when network violations are close to a previously existing edge and therefore approximate the learned network structure. When you learn something new that is similar to something you already know, you are only slightly surprised and quickly move on with your life. In contrast, when you learn something new that is unlike anything you already know, you often need time to think about it and integrate it with your existing picture of the world. This suggests that our brains learn **network distances**, or the spatial architecture of a network, not only between nodes that are directly connected, but also between distant nodes that are not connected in a network.

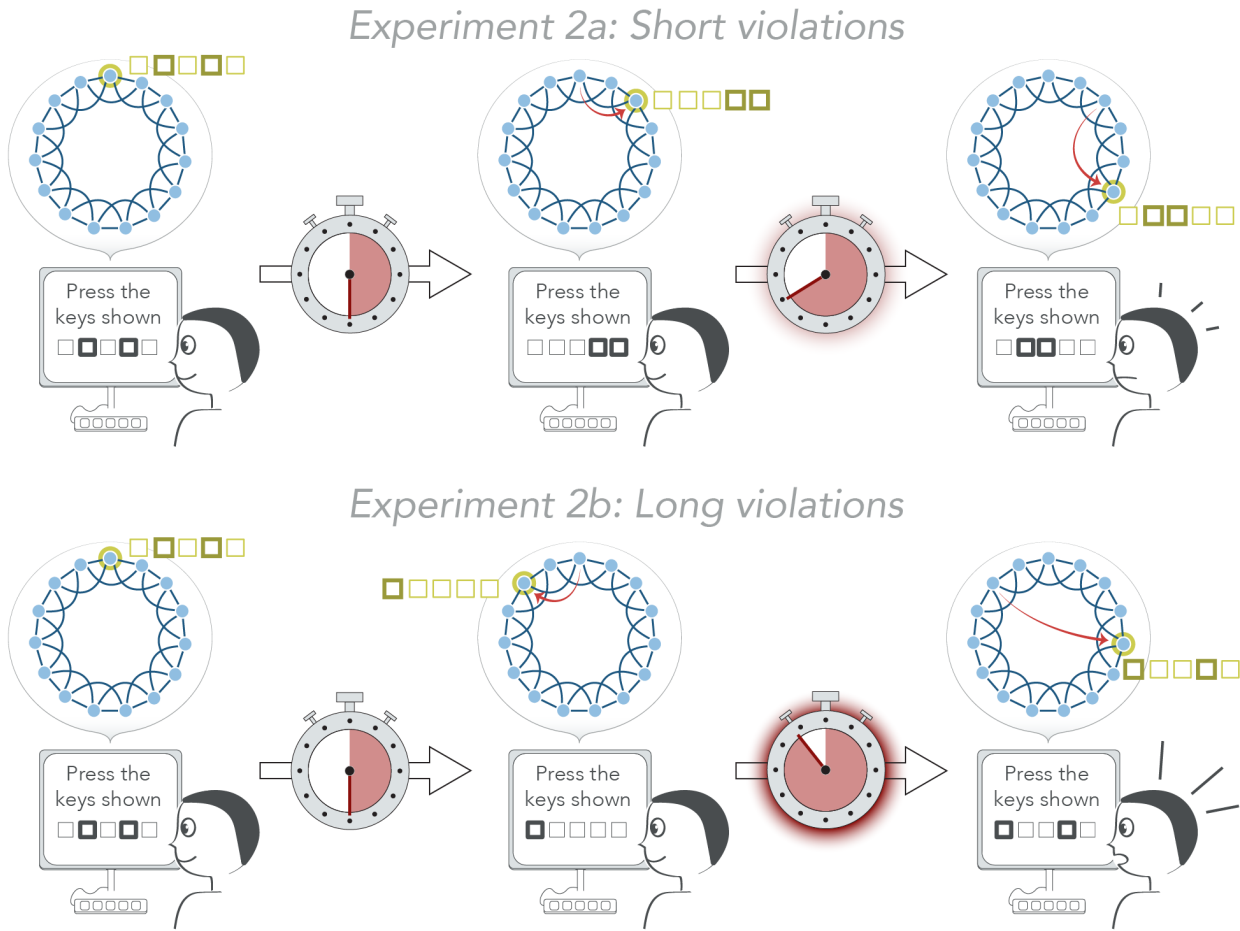

**Figure 3. Humans are surprised by ideas that violate their network model of the world.** When shown 5 squares, subjects clicked the corresponding keys on the keyboard. The images shown to subjects in this experiment were generated by transitions between nodes organized in a ring network (see Fig. 1b). When shown an incorrect transition of any kind, subjects took longer to respond. (b) When shown an incorrect transition that varied greatly from possible correct transitions or existing edges, subjects took longer to respond than (a) when shown an incorrect transition that remained close to existing edges.

## What can we do with this information?

Understanding *how* humans learn is necessary to further optimize modern teaching and communication methods. If we know that people learn information more easily when it is organized in a particular way, then we can make changes to the way information is presented, not only by teachers, but also in textbooks. For example, think of a history textbook. Usually, the information is ordered chronologically, but we know that modular networks are more easily learned. What if the lessons were instead grouped by similar themes or topics? Maybe all events having to do with politics could be in one section, while all events having to do with disease and public health could be in another section. Instead of having to remember facts with seemingly arbitrary connections, learners would be presented with a clear context, or network, in which to place new information. With these associations, perhaps the students could more efficiently learn the structure of what they read.

Organizing information for the purpose of teaching does not stop with textbooks. Children and adults are constantly reading, learning, and remembering, and some of these people have jobs where the information they learn has important and lasting effects on the world. Before making policy decisions, politicians are given reports written by scientists, which explain research findings and their relation to potential policy decisions. For example, scientists might explain that a policy prohibiting litter near a lake will keep local fish out of harm's way. We hope that future research will examine how the organization of information in scientific communication affects the policies that political leaders implement. We speculate that reports with information organized to demonstrate clear connections within categories, or organized in a "modular" structure, may improve the efficiency of communication.

## Contribution to the field

Past research on human learning has primarily focused on probing the way that people interact with individual pieces of information. While this first step is important, it does not accurately represent the way humans learn information in the world. In reality, humans are exposed to multiple pieces of information at once and must understand how these pieces are related. Network science enables human learning to be studied in a new way, taking into account not just how humans learn individual pieces of information, but also patterns and connections between information. This advance -- known as graph learning -- comes at an important time, as artificial intelligence is also being researched and developed. Both artificial intelligence and humans have the ability to learn patterns. By understanding better how humans learn and how our brains support that learning[5], not only are we able to optimize teaching for humans, we may also be able to optimize the development of artificial intelligence in similar ways. Conversely, understanding how artificial intelligence learns patterns could give insight into how humans learn patterns. Excitingly, our research in graph learning can open a line of communication between these two fields.

## Vocabulary Word Definitions
- **Node** - a thing in a network that can be connected to other things.
- **Edge** - a link that connects nodes in a network.
- **Degree** - the number of other nodes to which a given node is connected.
- **Transition** - the movement from one node to another connected node.
- **Network Structure** - a network's shape or what a network looks like, particularly its arrangement of nodes and edges.

# Author Biographies

**Sophia U. David:** Sophia David is a senior in High School at Friends' Central School, just outside of Philadelphia. She is currently working in the laboratory of Danielle Bassett at the University of Pennsylvania, creating network models to reflect how the brain takes in information and learns.

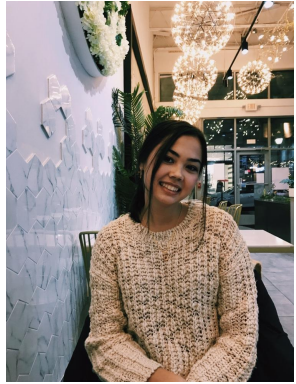

**Sophie E. Loman:** Sophie E. Loman is a first-year Ph.D. student in Dr. Danielle Bassett's Complex Systems Laboratory at the University of Pennsylvania. Loman received a B.S. in Cognitive and Brain Sciences from Tufts University, where they studied learning and behavior in planarian flatworms as a research assistant in Dr. Michael Levin's Developmental Biology and Regeneration Lab. Loman's current interests include network neuroscience, neuroimaging, graph learning, and psychiatry. As a nonbinary graduate student, Loman is committed to amplifying trans voices and advocating for trans inclusivity in academia.

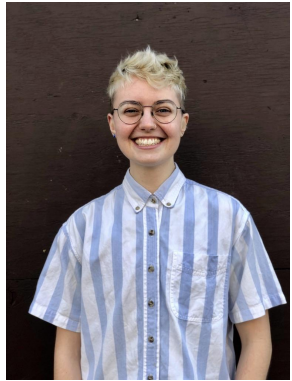

**Christopher W. Lynn:** Christopher W. Lynn is a postdoc in the Departments of Bioengineering and Physics & Astronomy at the University of Pennsylvania and a James S. McDonnell Postdoctoral Fellow at the Center for the Physics of Biological Function at the City University of New York and Princeton University. Lynn received a B.A. in physics and mathematics at Swarthmore College and a Ph.D. in physics at the University of Pennsylvania. Lynn's research combines ideas from information theory, network science, and cognitive science to study how humans learn and process information.

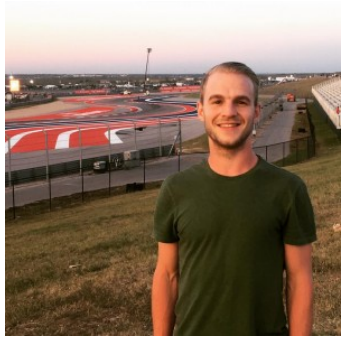

**Ann Sizemore Blevins:** Ann Sizemore Blevins is a postdoc in the Department of Bioengineering at the University of Pennsylvania. Blevins received a B.A. in mathematics and a B.S. in biology from Boston college as well as an M.S.E. and Ph.D. in bioengineering at the University of Pennsylvania. Blevins focuses on developing methods for biological systems using concepts from algebraic topology.

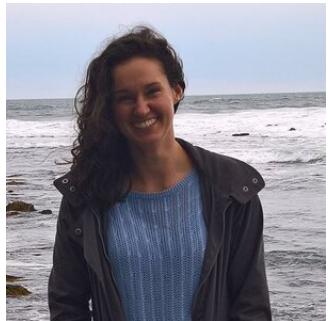

**Danielle S. Bassett:** Danielle S. Bassett is the J Peter Skirkanich Professor in the Departments of Bioengineering, Electrical & Systems Engineering, Physics & Astronomy, Neurology, and

Psychiatry at the University of Pennsylvania, and an external professor at the Santa Fe Institute. Bassett received a B.S. in physics from Pennsylvania State University and a Ph.D. in physics from the University of Cambridge, England, after which she spent several years as a postdoctoral scholar at the Sage Center for the Study of the Mind. Bassett teaches courses in mathematics, neuroscience, and curiosity and she is fascinated by networks, learning, and the nature of being human.

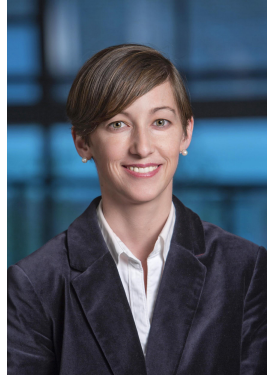